\documentclass[twocolumn,a4,showpacs,preprintnumbers,amsmath,amssymb]{revtex4}
\usepackage{epsfig}
\usepackage{graphicx}
\usepackage{dcolumn}
\usepackage{bm}

\begin{document}



\title { Shell model study of the pairing correlations} 

\author {J.A. Sheikh$^{1,2}$, P.A. Ganai$^{1}$, R.P. Singh$^{2}$, R.K. Bhowmik$^{2}$ and 
S. Frauendorf$^{3}$}
\address {$^1$Department of Physics, University of Kashmir, Srinagar,190 006, India \\
$^2$Inter-University Accelerator Center, New Delhi, 110 067, India \\
$^3$Department of Physics, University of Notre Dame,Notre Dame, IN 46556, USA
}

\begin{abstract}

A systematic study of the pairing correlations as a function of temperature
and angular momentum has been performed in the sd-shell region using the
spherical shell model approach. The pairing correlations have been
derived for even-even, even-odd and odd-odd systems near N=Z and 
also for the asymmetric case of N=Z+4. The results
indicate that the pairing content and the behavior of pair correlations
is similar in even-even and odd-mass nuclei. For odd-odd N=Z system, angular momentum
I=0 state is an isospin, t=1 neutron-proton paired configuration.  Further, these t=1 
correlations are shown to be dramatically reduced for the asymmetric case of N=Z+4.
The shell model results obtained are qualitatively explained within a simplified
degenerate model.

\end{abstract}


\maketitle

\section{Introduction}

It is well established that the pairing field is an important component
of the nuclear mean-field potential. The interplay between the deformation 
driving forces and the pairing field determines most of the properties of
a nuclear system. The relevance of the pairing field for the nuclear
many-body system was proposed in the pioneering work of Bohr,
Mottelson and Pines in 1958 \cite{bmp58}. There are two issues related to the
pairing potential which still need to be elucidated. The first issue
concerns with the detailed form of the pairing potential and the second
to the approximation employed for the solution of pairing force. Although, it is
quite evident from the analysis of the properties of nuclei, for instance the 
suppression of the moments of inertia of rotating nuclei and the observed
energy gaps, that pairing field is essential for describing 
atomic nuclei \cite{bm75,rs80}. But most of the properties of nuclei are rather insensitive
to the detailed form of the potential in the pairing or the particle-particle
channel. In comparison, most of the nuclear properties such as compressibility,
surface energy, effective masses, saturation and other specific properties
critically depend on the exact form of the nuclear potential in the Hartree Fock
or particle-hole channel. 
In addition, the potential in the particle-hole channel is constrained or
adjusted to the properties of the closed shell nuclei and the pairing 
field is introduced in an ad hoc manner in most of the density functional
theories (DFT). In most of these approaches, for instance
Skyrme and relativistic mean field models, the potential
is chosen to be of zero range and therefore, in principle, don't 
contain pairing correlations. In the finite range Gogny density functional
approach, the range of the effective pairing force is adjusted to the 
properties of $G$-matrix calculated from the bare nucleon-nucleon interaction. 

The second issue concerns with the method employed to solve the pairing interaction.
The pairing field has been primarily studied in the Bardeen-Cooper-Schrieffer (BCS) or 
Hartree-Fock-Bogoliubov (HFB) approximation.
In this approach, pairing field depicts a sudden transitional behavior as a 
function of rotational frequency and temperature \cite{good81,good88}. The pairing 
correlations are finite up to a certain rotational 
frequency or temperature and then these correlations suddenly vanish above
this transitional point. The empirical analysis of the experimental
data, however, does not depict this sudden phase transition and shows a smooth
transition from one phase to the other. The reason for this discrepancy
is known to arise from the neglect of the fluctuations in the mean-field
models \cite{er93,drr93,fkms03,aer01,srlr02}. 

It is known that relative size of the 
fluctuations becomes small for a system with a very large
number of particles, for example a bulk superconductor, and sudden
transitional behavior is a hall mark of these macroscopic systems. However,
for systems with finite number of particles, for instance atomic nuclei and
metallic clusters, the fluctuations are important and need to be incorporated
for an accurate description of these systems \cite{landau_stat,bfv99}. A 
powerful method to 
incorporate the fluctuations is the projection theory. In particular,
the particle number projection methods are now readily available to
incorporate the fluctuations quite accurately. However, these projection 
methods, before variation, are available only at zero 
temperature \cite{er821,er822,sr02,aer01,sdknt07} in the complete HFB framework. It needs to be 
added that the particle number projection analysis has been performed at finite
temperature with BCS ansatz \cite{ee93,nt06}. 

The projection at finite temperature is more important as it is known
that the mean-field or BCS wavefunction for an even or odd particle 
system has admixtures from both even and odd neighboring particle 
numbers \cite{bfv99,fo96}. At zero temperature, even (odd) system 
has admixtures 
from only even (odd) particle numbers. It has been also demonstrated
recently, in an exactly solvable model, that the pairing correlations
re-appear at finite temperature after they are quenched at zero temperature
and high rotational frequency \cite{fkms03,spf05}. This surprising result, 
which completely
contradicts the mean-field predictions, needs to be studied in more
realistic models. 

The spherical shell model (SSM) is another 
tool to study pairing correlations. The advantage of the SSM approach is
that, first of all, most of
the interactions employed in SSM are adjusted to open shell nuclei and,
 therefore,
contain a proper pairing force. Secondly, the exact diagonalization
of the Hamiltonian matrix is performed and results obtained contain
all the possible correlations including the pairing. There are many 
puzzling questions regarding the pairing correlations which can be addressed
within the framework of SSM. For instance, difference between the
pairing correlations in odd- and even-mass nuclei, importance of the
neutron-proton pairing correlations near N=Z and the phase transition
mentioned above. These issues originate when comparing the mean-field
results with the experimental data or exact solutions of toy models. It needs to
mentioned that in SSM approach only low-lying part of the excitation spectrum
is fitted and, therefore, the results obtained using SSM are inaccurate
for high-excitation energy or temperature.
 
The purpose of the present work is to study the pairing correlations
in a realistic space using the spherical shell model
approach. It needs to be emphasized here that a new shell model
program has been developed by two of the present authors \cite{ss06}. This new
program completely works in the j-representation \cite{fhmw69} and 
is quite similar in structure to that of NATHAN code developed by the 
Madrid-Strasbourg group \cite{cmnpz05}.
The shell model approach provides the most accurate description of nuclear 
properties and incorporates all the possible correlations. However,
the SSM analysis is restricted to lighter nuclei and its application to heavier
nuclei appears impossible in the near future. The most recent progress in the
SSM approach is the study of the fp-shell nuclei \cite{cmnpz05}.

In the present work, the shell model analysis has been performed in the
sd-shell region. The reason for choosing sd-shell is that, first of all, the
interaction is well established and it is known that ``USD'' interaction \cite{usd}
provides an accurate description of most of the sd-shell nuclei. Secondly,
it is required, in the present analysis, to calculate a few thousand 
states for each angular momentum
to evaluate the statistical partition function. These calculations
would become impossible in the fp-shell region. The calculations have
been performed in the middle of the sd-shell so that a large number of 
eigen-solutions are available for the statistical analysis. The nuclei for 
which the shell model calculations have been performed are :
$^{28}$Si, $^{27}$Si and $^{26}$Al. These three neighboring nuclei have been 
investigated so that a comparison can be made among even-even, even-odd and
odd-odd systems. The pairing correlations have also been deduced for the 
asymmetric system $^{24}$Ne in order to study the dependence of
the neutron-proton pairing on the particle number.

It is pertinent to mention here that while the
present work was in progress, a similar study of pairing correlations in
the sd-shell has been recently published \cite{hz07}. 
Although the model and the region of study
in the present work is same as that of ref. \cite{hz07}, the issues 
discussed are different and as a matter of fact the present work
complements the work of ref. \cite{hz07}. In the course of the discussion,
we shall comment on the results of ref. \cite{hz07} wherever necessary. 
The present manuscript is organized as follows : In the next section, the shell
model based expressions are presented for the evaluation of the pairing
correlations. The results of the calculations are presented and analyzed
in section III and finally the summary and conclusions are included
in section IV. 
  
\section{Shell model formulation}
The spherical shell model Hamiltonian, generally, contains single-particle 
and two-body parts and in the second quantized notation is written as
\begin{equation}
\hat H\,=\,\hat h_{sp}\,+\,\hat V_{2},  \label{E01}
\end{equation}
where, 
\begin{equation}
\hat h_{sp}\,=\,\sum_{rs} \, \epsilon_{rs} \,c_r^{\dagger}\,c_s,  \label{E02}
\end{equation}
and\\
\begin{eqnarray}
\hat V_{2}& = &{\frac{1}{4}} \sum_{rstu} <rs|v_a|tu> 
                 c_r^{\dagger}c_s^{\dagger}c_u c_t \nonumber \\ 
     & = & \sum_{rstu \Gamma}\,{\frac{\sqrt{(2\Gamma+1)}}
                    {\sqrt{(1+\delta_{rs})(1+\delta_{tu})}}}  
                    \,<rs|v_a|tu>_{\Gamma} \nonumber\\
                    &&\qquad \qquad
                      \biggl( A_{\Gamma}^{\dagger }(rs)
                         \times \tilde A_{\Gamma}(tu)\biggr)_{0},
\label{E03}
\end{eqnarray}
where $\epsilon_{rs}$ are the single-particle energies 
of the spherical shell model
states, which are diagonal except in the radial quantum numbers and 
$<rs|v_a|tu>_\Gamma$ are the two-body interaction matrix elements and in the
present work are chosen to be those of ``USD''. The two-particle 
coupled operator in Eq. (3)
is given by $A_{\Gamma}^{\dagger }(rs)=
(c_{r}^{\dagger }c_{s}^{\dagger })_{\Gamma}$ and $%
\tilde A_{\Gamma M_\Gamma} = (-1)^{\Gamma - M_\Gamma} A_{\Gamma -M_\Gamma}$. The labels $r,s,...$
in the above equations denote the quantum numbers of spin and isospin. 
``$\Gamma$'' quantum number labels both angular momentum and isospin
of the coupled state. The above notation is same as that used in 
ref. \cite{fhmw69}.

In the present work, the pairing correlations have been calculated
using the canonical ensemble approach since the exact 
solutions have well defined
particle number. The average value of a physical quantity ``F'' in canonical
ensemble is given by \cite{fkms03,landau_stat}
\begin{equation}
<<F>> = \sum_{i} F_i e^{-E_i/kT}/Z,
\end{equation}
where,
\begin{eqnarray}
Z     &=& \sum_{i} e^{-E_i/kT} \\
\hat H |i> &=& E_i |i> \nonumber \\
F_i   &=& <i|\hat F|i>
\end{eqnarray}
In the partition function, Eq. (5), $k$ is the Boltzman
constant and $T$ is the temperature. In the rest of the manuscript,
we shall use ``Temp'' rather than $kT$, which has dimensions
of energy (MeV ) and the symbol $T$ shall be used for the isospin
of the coupled two-particle state.

In the present work, we study the isovector monopole pair correlations 
for the ``$\Gamma_0=\{I=0,T=1,|T_z|=0,1\}$''. The canonical
average for this is calculated as
\begin{eqnarray}
E_{\Gamma_0}(\rm pair) &=& \sum_{rstu}\,{\frac{\sqrt{(2\Gamma+1)}}
                    {\sqrt{(1+\delta_{rs})(1+\delta_{tu})}}}  
                    \,<rs|v_a|tu>_{\Gamma_0} \nonumber \\
&& \qquad << 
(A_{\Gamma_0}^{\dagger } \times \tilde {A}_{\Gamma_0})_{0}>>,
\end{eqnarray}
from the energies and eigen-states obtained by diagonalization,
which shall be referred to as the ''correlated'' pairing energy 
$E_{pair}(corr)$.
We have also calculated the ``uncorrelated'' pairing energy by removing
the monopole field $\Gamma_0$ in the interaction and 
then redoing the shell model diagonalization \cite{fkms03}.
Using the resulting energies and wavefunctions, Eq. (7)
is evaluated to give the ``uncorrelated'' pairing energy.
 It is 
noted that the definition of present pairing energy is slightly different from that 
used in refs. \cite{hz07,dean}. In
these studies, it is calculated without the matrix elements
and the constant factors in front of the expectation values in Eq. (7).

\section{Results and discussions}

The shell model calculations have been performed in the middle of the sd-shell
for $^{28}$Si (with six valence protons and six valence neutrons), $^{27}$Si
(with six valence protons and five valence neutrons) and $^{26}$Al (with
five valence protons and five valence neutrons). 
The calculations have also been
performed for $^{24}$Ne (with two protons and six neutrons)
in order investigate the behavior of the pairing correlations with asymmetry
in proton and neutron particle numbers. The temperature and the
angular momentum dependence of the pair correlations have been
studied in detail for these systems and are discussed in subsections B and C.
In subsection D, the temperature dependence of the average isospin is discussed.
Before presenting these
results, in subsection A a simple model is briefly described to analyze
the essential results obtained with the full shell model calculations.

\subsection{Isospin geometry and a simple model}

Before starting the discussion, we first briefly recall some consequences of
isospin being a good quantum number. Writing explicitly the isospin and omitting the
other quantum numbers, i.e. $\Gamma=T,T_z$, 
the pair energy is measured by 
\begin{equation}
A_{1,M}^{\dagger } \times \tilde {A}_{1,-M}=\sum_{t=0,1,2}<1M1-M|t0>B_{t0},
\end{equation} 
where the proton-neutron paring corresponds to $M=0$ and the neutron-neutron pairing to $M=1$.
Since $A_{1,M}^{\dagger }$ and $\tilde {A}_{1,-M}$ are tensor operators
in isospace their product can be rewritten as a sum of tensor 
operators $B_{t0}$ and 
\begin{eqnarray}\label{decompo}
<TT_z|A_{1,M}^{\dagger } &\times& \tilde {A}_{1,-M}|TT_z>\nonumber \\
&=&\sum_{t=0,1,2}<1M1-M|t0> \nonumber \\
&\times&<TT_z10|TT_Z><T||B_{t}||T>.
\end{eqnarray} 
Only the $t=0$ term contributes in $T=0$ states, which means
that $<T=0|A_{1,M}^{\dagger } \times \tilde {A}_{1,-M}|T=0>$ does not depend
on $M$, i. e. $E_{pair}$ is the same for all three types of pairing.  
For $T>0$ states, the terms $t=1,2$ contribute as well, which means  
 $E_{pair}$ may be different for each pairing channel.

For a quantitative
statement, one needs to know the reduced matrix elements
$<T||B_{t}||T>$, which depend on the detailed structure of the shell model states.
For the lowest states, pairing correlations are strong.
In such a case one expects that the model of isovector monopole pairing 
in a degenerate shell should allow us some rough estimate of the relative 
paring strengths. The model is discussed in \cite{engel96,engel98}, where the 
original work is cited and explicit expressions 
are given in terms of number of nucleon
pairs ${\cal N}$ in the shell, the isospin $T,T_Z$, and the number of pairs 
$\Omega(=12)$ that the shell can accommodate, where the 
unit is the coupling constant
$G$. It was found there that the model accounts well for
relative strengths of the pair energies but cannot reproduce the scale
of the shell model calculations, which was attributed to the splitting 
of the levels due to the deformation and the spin-orbit coupling.     
Following this observation, we scale the pair energies of the degenerate
model by a common factor. 

\subsection{Temperature dependence of the pair correlations}

The results of the neutron-neutron 
and neutron-proton monopole pair
correlation energies for $^{28}_{14}$Si$_{14}$ are 
shown as a function of temperature (Temp) 
in Figs. 1 and 2 for even- and odd-spin values. For the symmetric system
and isospin invariant two-body interaction, the proton-proton pairing
energy is identical to that of neutron-neutron.
The reason is that the canonical ensemble contains mainly isospin, $T=0$ states, which  
lie lower than the $T>0$ states in the N=Z=even nuclei \cite{jaenecke02}. 

Even- and odd-spin
values are plotted separately in two figures since they have different intrinsic 
structures. In the
quasi-particle language, the low-lying even-spin members in an even-even
nucleus have 0-quasiparticle intrinsic structure and the odd-spin members
have 2-quasi-particle structure. Obviously, the fraction
of $T>0$ states is larger for odd spin. 

In order to investigate, in detail, 
the variation of the pairing correlations with spin ($I$), the correlations
are plotted separately for each possible value of spin and are shown in 
two  panels. In the top panel the pairing
correlations are calculated with full two-body interaction [denoted
by $E_{pair}(corr)$]. The lower panel
depicts pairing correlations [denoted by $E_{pair}(uncorr)$] in which the 
monopole terms in the two-body
interaction were excluded.

For $I=0$ in Fig. 1, the $E_{pair}(corr)$ are quite similar for neutron-proton
and identical particle channels and are also quite large 
at low temperatures. For Temp=0, they are  \mbox{4.6 MeV}. The 
ground-state in even-even systems is a paired configuration with 
maximum correlations. The degenerate model gives 3/2 for the correlation
energy, which fixes the scale factor to \mbox{3.06 MeV}. In the following
we shall use this factor
to scale other calculations in the framework of the degenerate model.  

The pair correlations are almost 
unchanged till  Temp$\simeq$2 MeV and then are observed to 
be reduced with
increasing temperature. However, in comparison to the mean-field models
which predict a sudden transition from the paired to the unpaired state,
the exact analysis depicts a smooth drop in the pair correlations. The phase
transition obtained at Temp$\simeq$2 MeV in Fig. 1 is higher as compared to 
the transition point obtained in HFB and also in shell model Monte-Carlo study,
which predict at Temp$\simeq$0.5 to 0.7 MeV \cite{lan06}. It is to be noted that
in the present work, the pairing correlations have been obtained using the expression,
Eq. (7) and, as already pointed out at the end of section II, is slightly different from 
the expression used
in the HFB and the shell model Monte-Carlo studies. In order to confirm that
the reason for the
discrepancy in the phase transition point is due to the different pairing expressions
used, we have 
performed shell model calculations for a simpler case of $^{24}$Mg  with the 
pairing expression as used in
HFB and Monte-Carlo studies and the phase transition was found to be around 0.9 MeV. It 
is also noted 
that the magnitude of the pairing energy 
obtained in the present study is
also quite large as compared to the HFB and the shell model Monte-Carlo results.
Further, it is seen in Fig. 1 that at large temperatures
the neutron-proton pairing energy deviates from proton
and neutron pairing energies, which was also observed in 
ref. \cite{dean}. The difference between the proton-neutron and 
neutron-neutron (=proton-proton) 
pair energies indicates some admixture of $T=1$ states to the ensemble.

For even-spin values of $I$=2, 4 and 6,  it is observed that the pair
correlations drop in a step wise manner and the pairing gaps for these spin 
values at Temp=0 are approximately 
3.9, 3.2 and 2.6. The 
temperature dependence of
the pair energies for $I=2$ and 4 show a similar behavior as that of $I=0$.
For I=6, the pair correlations are almost constant with temperature. 
The shell model calculations for $I=8$ and 10 (not shown in the figure) 
depict a slight increase in the pair correlations  at low temperatures
and for higher temperatures the 
correlations drop as for the earlier cases.

The uncorrelated pairing energy are shown in the lower panels of Fig. 1.
As already mentioned, they have been obtained by a second  shell
model diagonalization on setting the monopole matrix elements equal to zero.
The problem in the shell model study of the pair correlations is that
the calculated correlated pairing energy may contain the contributions
from other multipoles due to recoupling, which in the mean-field language
are referred to as the particle-hole contribution. In the mean-field 
framework, particle-particle and the particle-hole 
channels are completely decoupled
and the pairing energy can be directly evaluated from the pairing potential.

It is noticed from Fig. 1 that
the uncorrelated pairing energy  is substantially smaller as compared to
the correlated one.
However, it shows a similar transitional behavior 
with temperature as the correlated pairing energy. In order to explore
the reason for this unexpected transitional behavior in the uncorrelated
energy, we have also removed I=1 and 2 apart from I=0 matrix elements
in the shell model diagonalization and the pairing energy again calculated
using the expression of Eq. (7). The results indicate that the phase transition
is now almost washed out in the uncorrelated pairing energy. 
We are presently performing a detailed investigation of this phenomenon 
and the results of this study
 will be presented elsewhere.

The neutron-proton uncorrelated pairing energy 
appears to be lower than
that of the corresponding neutron and proton energies. This can be
understood as follows. The isovector monopole
pair correlations shift the $T=0$ states to lower energy as compared
to the $T>0$ states. If these correlations are switched off, then
this preference is lifted. The increased fraction of $T>0$ states
in the ensembles creates the difference between the neutron-neutron and 
proton-neutron pairing.
However, using this simple perspective, it is not possible to understand
why the like-particle pairing is stronger.


The odd-spin values are depicted  in Fig. 2.  The pair energies 
are smaller than  the even-spin values at very 
low temperature. This is easily understood by noting that the 
odd-spin band have two
quasi-particle structure with reduced pairing. For $I=1$, 
the yrast state has $T=0$.
The pronounced increase of the neutron-proton pair energy as compared to the like-particle 
ones indicates that low-lying $T>0$ states become a substantial fraction
of the ensemble. 

For the larger angular momenta, $I=7$ it is observed that
the pair correlations first {\it increase} with temperature. 
It is well known that in the mean-field theory, the pair correlations
are reduced with both increasing temperature and rotational frequency. The
mean-field analysis always depicts a phase transition from the paired to 
the unpaired phase with increasing temperature and rotational frequency.
This is experimentally observed in macroscopic  
superconductors, where the
particle number is quite large. However, for mesoscopic systems 
the mean-field solution is inappropriate.
The exact solutions of the pairing problem do not have a
sharp phase transitions. Further, they may have peculiarities, as e.g.,
recently demonstrated in a simple few-level 
model that the pairing correlations re-appear at finite temperature 
after they are quenched at zero temperature and high rotational frequency
\cite{fkms03}, where this peculiar effect is explained. 
The shell model results for $I=$9 and 11 (not shown in the figure) also 
depict a minor increase.
Therefore, increase in the pairing correlations with increasing temperature
obtained in the earlier work is an artifact of the simple model
employed \cite{fkms03} as the present results in the sd-shell don't depict
such a dramatic rise. However, it needs to be added that in the sd-shell,
it is not possible to have a large angular momentum where the 
re-appearance of the pairing correlations is predicted in the earlier
simple model study.

The results of the pair energies for the odd-system $^{27}_{14}$Si$_{13}$ 
are presented in
Fig. 3. The results presented in this figure also  
correspond to $^{27}_{13}$Al$_{14}$ with proton and neutron curves interchanged.
At zero temperature the expected picture evolves. The proton-proton energy of 
\mbox{4.4 MeV} is similar to the value of 4.6 MeV in the even-even 
neighbor $^{28}_{14}$Si$_{14}$, as both have same proton number.
The neutron-neutron pair energy of \mbox{3.0 MeV} is about 30\% lower due to the blocking
of one level in the odd-neutron system. This is in accordance with   
the mean-field predictions that the pair gaps
in an odd-system is reduced by about $15\%$ in comparison to the 
neighboring even-even system, which amounts to a reduction of about 30\% in the 
correlation energy. The np- pair energy of \mbox{3.7 MeV} is less
reduced than the neutron-neutron one. This is understood, because the odd neutron
blocks the level only partially for proton-neutron pairs. [If the odd neutron is 
in the state 
$m>0$, a pair with the neutron in $m<0$ and 
the proton in $m>0$ may by accommodated.]
The  pair energies are different, which is expected
for  $T\leq 1/2$ states from Eq. (\ref{decompo}). The degenerate model
allows us to estimate the consequences of blocking. For $T= 1/2, ~{\cal N}=4$ 
it gives 
$E_{pp}(corr)=3/2,~E_{np}(corr)=15/12,~E_{nn}(corr)=1$. The respective
scaled values of 4.6, 3.8 and 3.1 MeV compare well with the shell model values
of 4.4, 3.7 and 3.0 MeV.
 
In early nineties, it was 
demonstrated through a series of systematic studies \cite{bgws92} 
of experimental data 
that the difference between the moments of inertia of neighboring 
even-even and odd-nuclei is merely $\leq 2\%$. This posed a serious 
challenge to the traditional nuclear structure models based on mean-field
theory which give a difference of about $15\%$. It was shown later using
number projected mean-field models that this difference in the moments of
inertia could be reduced. The moment of inertia depends on deformation and
the pairing correlations and considering that
$^{28}$Si and $^{27 }$Si have similar deformation values, it is expected 
that the moments of inertia of two nuclei would be similar since the pair gaps 
of the two systems are similar.The calculated moments of inertia 
for low-lying states for $^{28}$Si and $^{27}$Si are 2.62205 $\hbar^2/(MeV)$ 
and 2.66220 $\hbar^2/(MeV)$, respectively. 

The results for the odd-odd $^{26}_{13}$Al$_{13}$ system 
are presented in Figs. 4. For low-spin, the
pair energies are rather different
from those of the even-even and odd-systems. For $I=0$, 
neutron-proton pairing energy is quite large as compared to like-particle
pairing energies.
The ground states in odd-odd self-conjugate nuclei show a preference for
$T=1$ \cite{jaenecke02} with a $T=0$ state close by. 
In our case the ground state has $T=1$. 
For $T=1,~T_z=0$, the simple model of isovector pairing
in a degenerate shell gives a ratio 
$E_{pn}(corr)=43/15,~~E_{nn}(corr)=E_{pp}(corr)=11/15$. The respective
scaled values of 7.8 and 2.2 MeV are to be compared with the shell model
results of 7.9 and 2.7 MeV. The proton-neutron pair energy remains larger than 
the neutron-neutron pair energy for finite temperatures, which indicates that the 
isovector proton-neutron pair correlations lower the $T=1,~T_z=0$ states relative
to the $T=0,~T_z=0$ states.  
The lowest $I=2$ state has $T=0$ and the same pairing 
energies for like and unlike 
particles. For higher temperature,
neutron-proton pairing is  larger as compared to identical particle pairing,
which indicates a large fraction of $T=1$ states in the ensemble.
With increasing spin the results become similar to the even-even system,
which can be understood as a consequence of the quenching of the
pair correlations in both types of nuclei. This is corroborated
by the observation  for the odd-A case that the different pair energies become
similar (though not equal) at large spin. 
For odd-spin values, the pair energies are nearly equal for
all spin values. Only  at higher temperatures there is a slight asymmetry.
This indicates that the odd-spin states have preferentially $T=0$.
 
The pair correlations for the asymmetric
system $^{24}_{10}$Ne$_{14}$  are presented in Fig. 5.
As expected,  the neutron pair energy is larger than 
the proton one, because there are more neutrons in the open shell. 
The neutron-proton pair energy is  quite small.
In contrast to the symmetric even-even system, where it is as large  as the 
like particle one. This is in accordance
with  the HFB theory with neutron-proton pairing \cite{goodman72}, which finds 
quite strong neutron-proton pairing   for the N=Z system, but vanishing
for the asymmetric case of N=Z+4. The present shell model substantiate
these HFB results, although the neutron-proton has not vanished in the shell
model study but has clearly become quite weak. The reason is that the 
extra neutron pair blocks a level for the proton-neutron 
pairs to scatter into.
For zero temperature, the difference between 
the symmetric and asymmetric systems is qualitatively
reproduced by the simple model of a degenerate shell. 
The ground state has ${\cal N}=4,~T=2$,  which gives 
$E_{pp}(corr)=13/14,~~E_{nn}(corr)=27/14,~~E_{np}(corr)=13/42$.
The respective scaled values are 2.8, 5.9 and 1.0 MeV to be compared with 
the shell model results of 3.4, 5.5 and 1.2 MeV.

The model of a degenerate shell obeys the dynamical
symmetry of the group SO(5) with the isospin subgroup SU(2).
As will be discussed in a forthcoming paper, the relative
strength of the three types of isovector pairing 
mainly reflects the geometry of the isospin induced by SU(2),
which remains valid for the case of a non-degenerate shell.
The full SO(5) symmetry is exploited when the effects of blocking
in states with non-zero seniority  are estimated as for 
the case of odd-A in this paper. 

\subsection{Angular momentum dependence of the pair correlations}
The angular momentum dependence of the pair correlations for the four
systems studied are depicted in Fig. 6. The results for $^{28}$Si
indicate that for zero temperature, 
the pair correlations drop monotonically
with increasing angular momentum, except for $I=1$ which shows a larger drop.
In the case of $^{27}$Si, the pair correlations show
a staggering effect,
where the phase of the staggering is same for the three
pairing modes.

For $^{26}$Al, neutron-proton correlations are maximal for $I=0$. This 
is evident from Fig. 6 with a drop of about 5 MeV from $I=0$ to 1 in the correlated
pairing energy. For the higher spin values, it is noted that pair correlations
drop very little.  
The results for $^{24}$Ne depict a larger staggering effect as compared to
$^{28}$Si for identical particle pairing. The neutron-proton pairing shows
an irregular behavior with spin for this asymmetric system. 
As expected, for higher 
temperature of 3 MeV, the pair correlations depict a smoother behavior
with spin.

\subsection{Isospin analysis}

In most of the analysis presented in the above subsections, the isospin
content of the states played a crucial role to understand the behavior of the
pairing energies. It is, therefore, imperative to ascertain the temperature
dependence of the isospin. It needs to be mentioned that the new shell model program
developed \cite{ss06} uses the neutron-proton product basis and, therefore,
isospin, although conserved, needs to be evaluated for each eigen-state. The
expectation value of $\hat T^2$ has been calculated using the shell model
wavefunctions as discussed in ref. \cite{kbo78}. The average value of $T^2$ 
as a function of temperature has been obtained by using the canonical
partition function in the same manner as the pairing correlations have
been deduced.

The average value of $T^2$ is shown in Fig. 7 for the lowest angular momentum
ensemble. In the top panel of Fig. 7,
average value of $T^2$ is plotted for $^{28}$Si and it is quite evident from this
figure that the isospin at low temperature is T=0 up to Temp=2 MeV. Above this
temperature, the isospin increases steadily and correlates well with the drop in
the pair energies observed in Fig. 1. For the odd-odd $^{26}$Al system, the isospin
at low temperature is equal to one and giving rise to large difference in the
pairing correlations between neutron-proton and identical particle channels in
Fig. 4. However, with increasing temperature, it is noted that the average isospin
drops as T=0 states enter into the ensemble with the consequence that neutron-proton
pairing energy comes closer to the identical particle pairing at higher temperatures
in Fig. 4. 

For the asymmetric case of $^{24}$Ne, the average isospin in Fig. 7 
is equal to two and appears to be almost constant with increasing temperature. This
constancy of the average isospin gives rise to the constant behavior of the
neutron-proton pairing energy in Fig. 5. The temperature behavior of the
isospin for the odd-mass system $^{27}$Si is similar to the even-even system 
$^{28}$Si and is the reason that the behavior of the pairing correlations for
the two systems are quite similar.

\section{Summary and conclusions}

In the present work, the shell model study of the pairing correlations has
been undertaken. 
The calculations have been performed in the
sd-shell for $^{28}$Si, $^{27}$Si, $^{26}$Al,
and $^{24}$Ne. For the case of the even-even 
system $^{28}$Si, the pair energy of even-spin states 
as a function of temperature depicts a smooth but pronounced 
decrease around Temp=2 MeV,
which can be interpreted as the strongly 
washed out relic of the phase transition.
In the case of the odd-spin states up to $I=7$, the pairing 
correlations decrease only very slowly with increasing temperature, starting
from a reduced value, which  is caused by the blocking of two levels
in the states carrying two quasi particle character.

It is also clearly evident from the present study that pairing correlations are
non-zero  even at large temperatures and angular momenta. This is in contradiction
to the mean-field predictions that pairing correlations die out at higher temperatures
and angular momenta. For the very small systems studied, most of the pair correlations 
is generated by the fluctuations 
of the pair field, which are more prominent than the mean-field itself.

The proton-proton, neutron-neutron, and proton-neutron isovector pair correlation energies are not 
scalar under rotation in isospace, which means that they are only
equal in $T=0$ states but generally different in $T>0$ states.    
For the ground states of even-A nuclei, the relative strengths of the different
pairing energies is qualitatively reproduced by the simple model of a degenerate shell.
The even-even $N=Z$ nucleus has a $T=$0 ground state and thus equal proton-proton, neutron-neutron,
and proton-neutron strength. The proton-neutron strength deviates from the identical ones with
increasing temperature because $T>0$ states enter the ensemble.
The odd-odd $N=Z$ nucleus $^{26}_{13}$Al$_{13}$ has a $T=1,~T_z=0$ ground state for which
the proton-neutron strength is about three times larger than the like particle strength.
With increasing temperature, $T=0$ states enter the ensemble, reducing the 
the proton-neutron contribution. The asymmetric nucleus  $^{24}_{10}$Ne$_{14}$ has a $T=2, T_z=2$
ground state with the proton-neutron pairing dramatically reduced as compared to the symmetric 
systems $^{28}$Si and $^{26}$Al. In the odd-neutron nucleus $^{27}_{14}$Si$_{13}$,
the proton-proton strength is about the same as in $^{28}_{14}$Si$_{14}$, the neutron-neutron strength is
about 30\% lower and the proton-neutron strength in between, which reflects the blocking of
one level by the odd neutron.  

Finally, we would like to mention that the results of the present work are questionable at
higher temperatures as the configuration space of sd-shell employed in the present work is
suited only for low-excitation energies. For higher temperatures, it is expected that fp-shell
will be populated and for accurate evaluation of pairing correlations it is essential 
to include fp-shell configuration space in the shell model analysis. However, it is impossible
to perform shell model calculations with a complete sdfp-configuration space. What is feasible
is to calculate the partition function of the fp-shell in a spherical degenerate limit 
and then calculate
the total partition function with the method illustrated in ref.\cite{hm72}. We are presently 
working to evaluate the pairing correlations using this approach and the results of this
analysis shall be presented in the near future.

\newpage

\begin{figure*}
\includegraphics {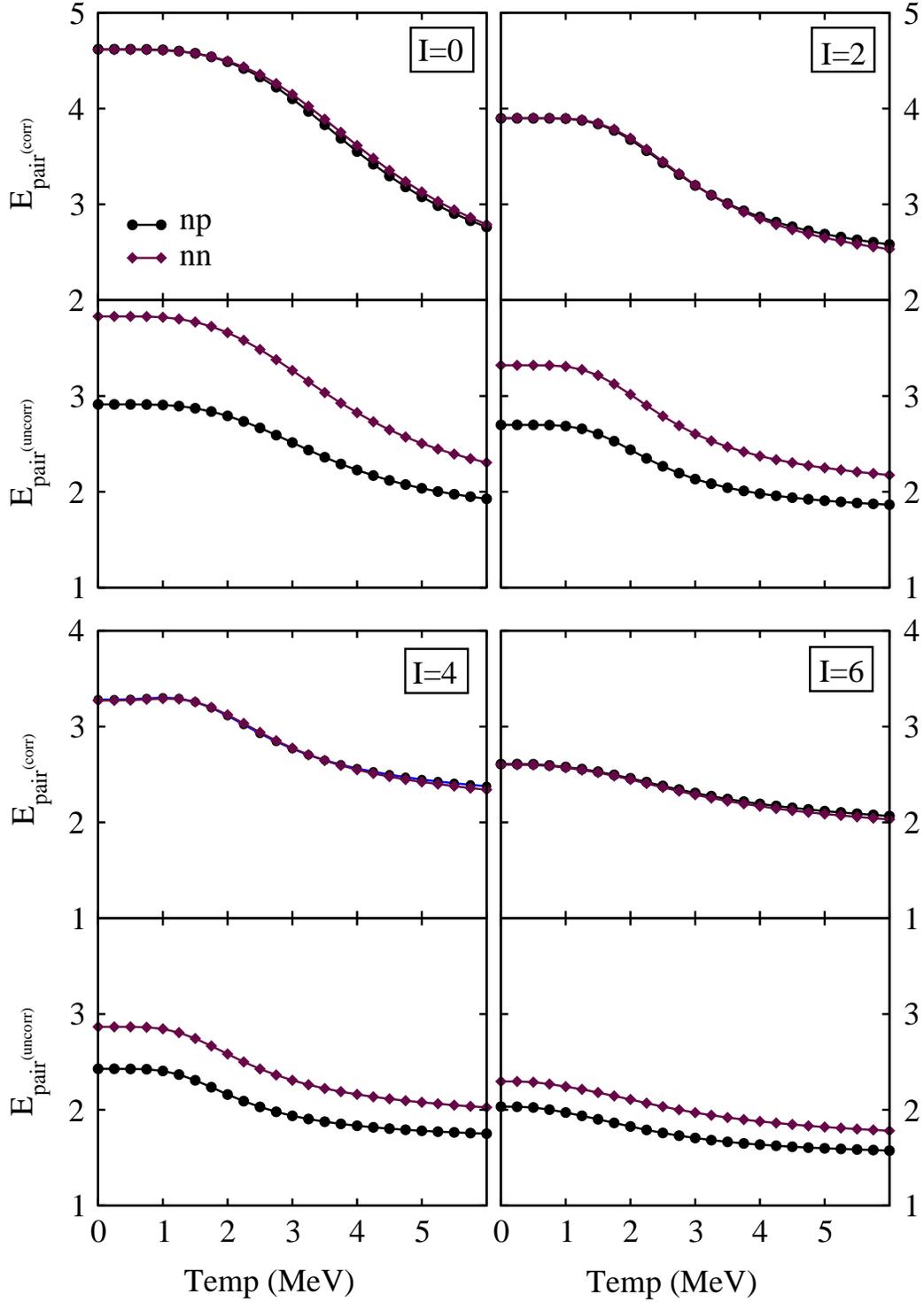}
\caption{(color online) Temperature dependence of pair correlations for $^{28}$Si. The results are shown for even-spin values of I = 0, 2, 4, 6. For each spin
state, there are two panels. The upper panel depicts the pairing correlations
calculated with the full two-body interaction. The lower panel shows the 
pairing energy without monopole matrix elements, referred to as the uncorrelated contribution. For 
I=0, the  temperatures 
of 1,2, 3, 4 and 5 MeV correspond to excitation energies of 0.04, 0.90, 4.0, 9.57 and
15.38 MeV, respectively. 
}
\label{figure.1}
\end{figure*}

\begin{figure*}
\includegraphics {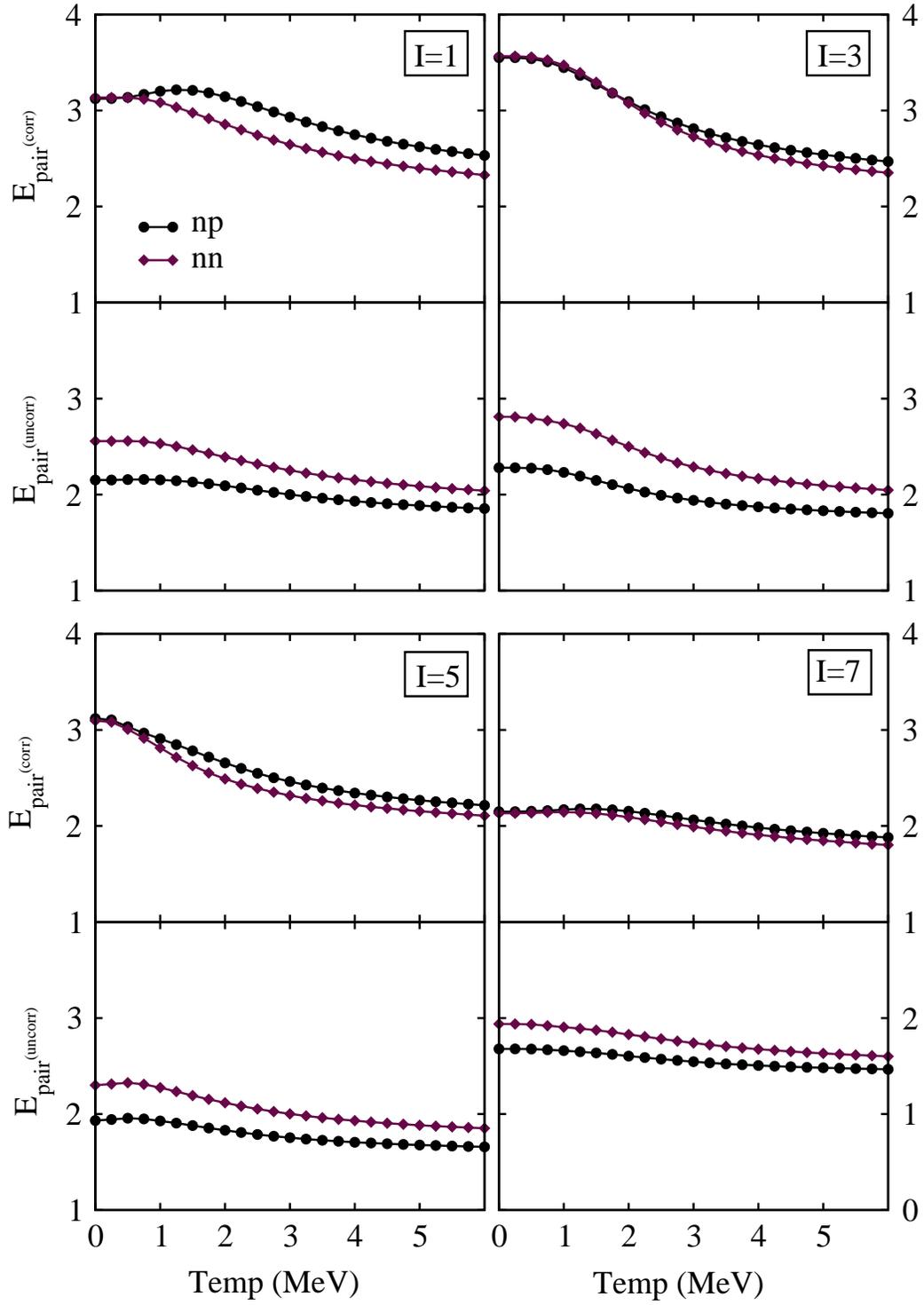}
\caption{(color online) Temperature dependence of pairing energy  for $^{28}$Si. Pair energy
is shown for odd-spin values with I = 1, 3, 5 and 7.
}
\label{figure.2}
\end{figure*}

\begin{figure*}
\includegraphics {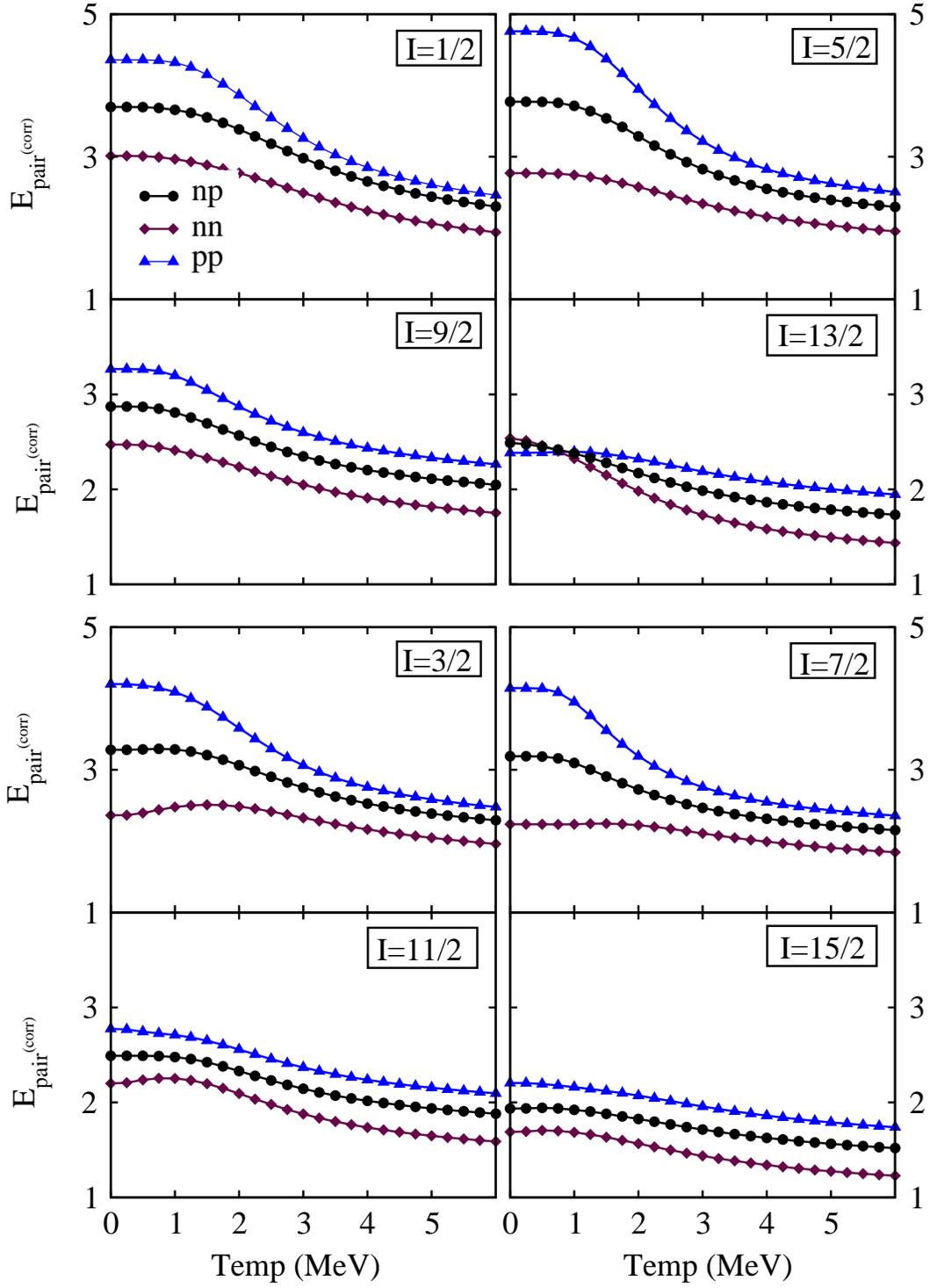}
\caption{(color online)Temperature dependence of pairing energy for $^{27}$Si. The results are
shown for I = 1/2, 5/2, 9/2 and 13/2 on the upper part and for I = 3/2, 7/2, 11/2 and 15/2 on bottom panels.
}
\label{figure.3}
\end{figure*}

\begin{figure*}
\includegraphics {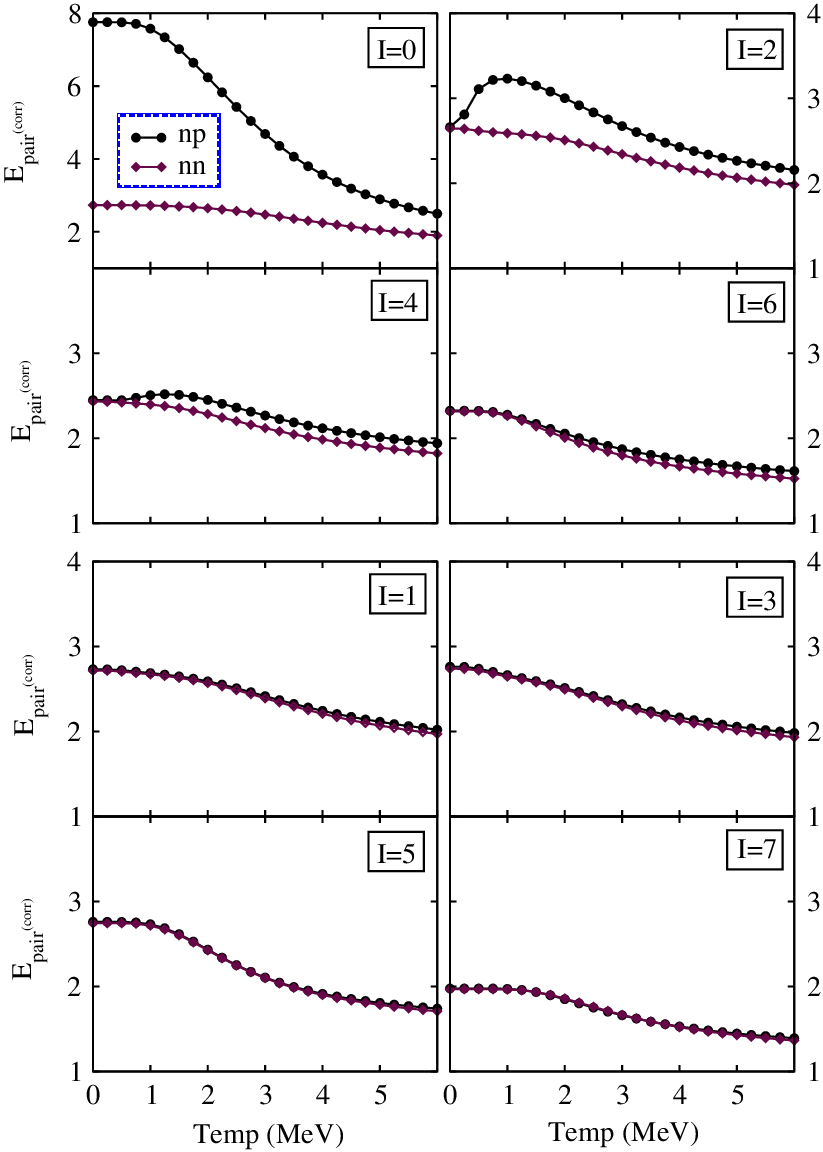}
\caption{(color online) Temperature dependence of pairing energy for $^{26}$Al. The pair-gaps
are plotted for I = 0, 2, 4,and 6 on upper two panels and  for I = 1, 3, 5,and 7 on the lower two panels.
}
\label{figure.4}
\end{figure*}

\begin{figure*}
\includegraphics {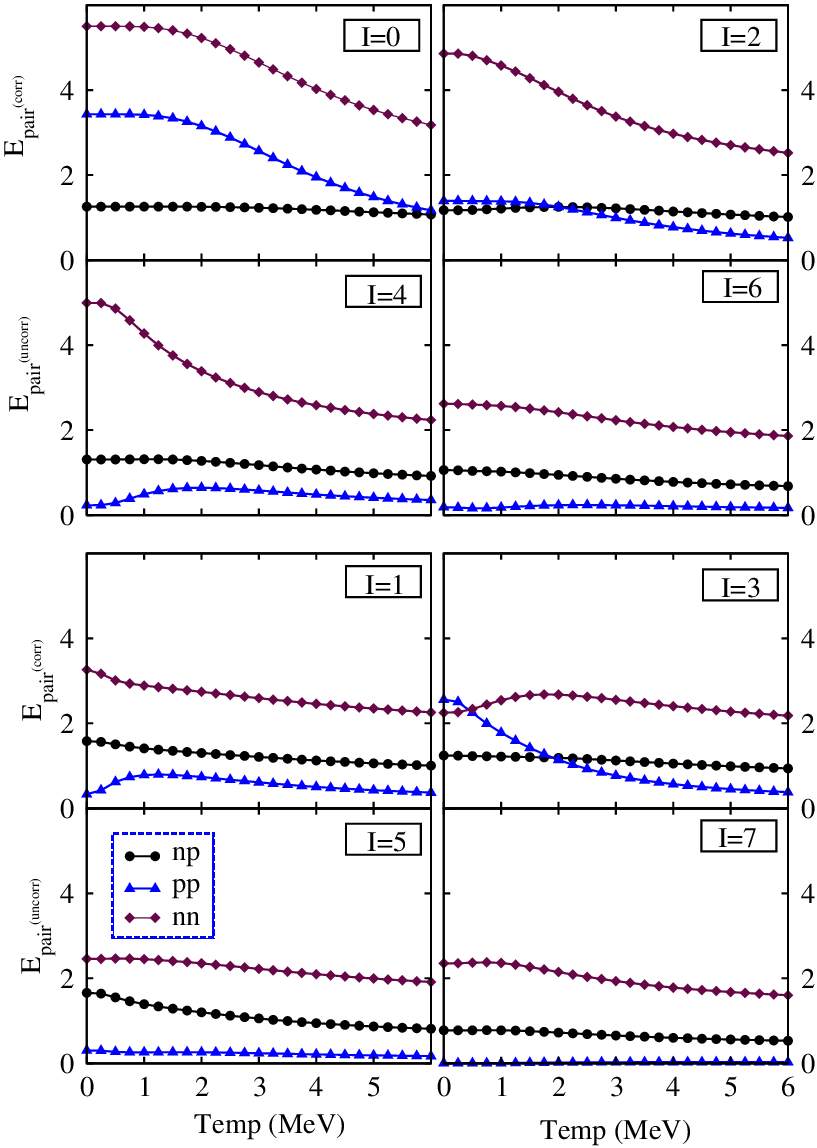}
\caption{(color online)Temperature dependence of pairing energy for $^{24}$Ne.The pair-gaps
are plotted for I = 0, 2, 4,and 6 on upper two panels and  for I = 1, 3, 5,and 7 on the lower two panels.}
\label{figure.5}

\end{figure*}

\begin{figure*}
\includegraphics {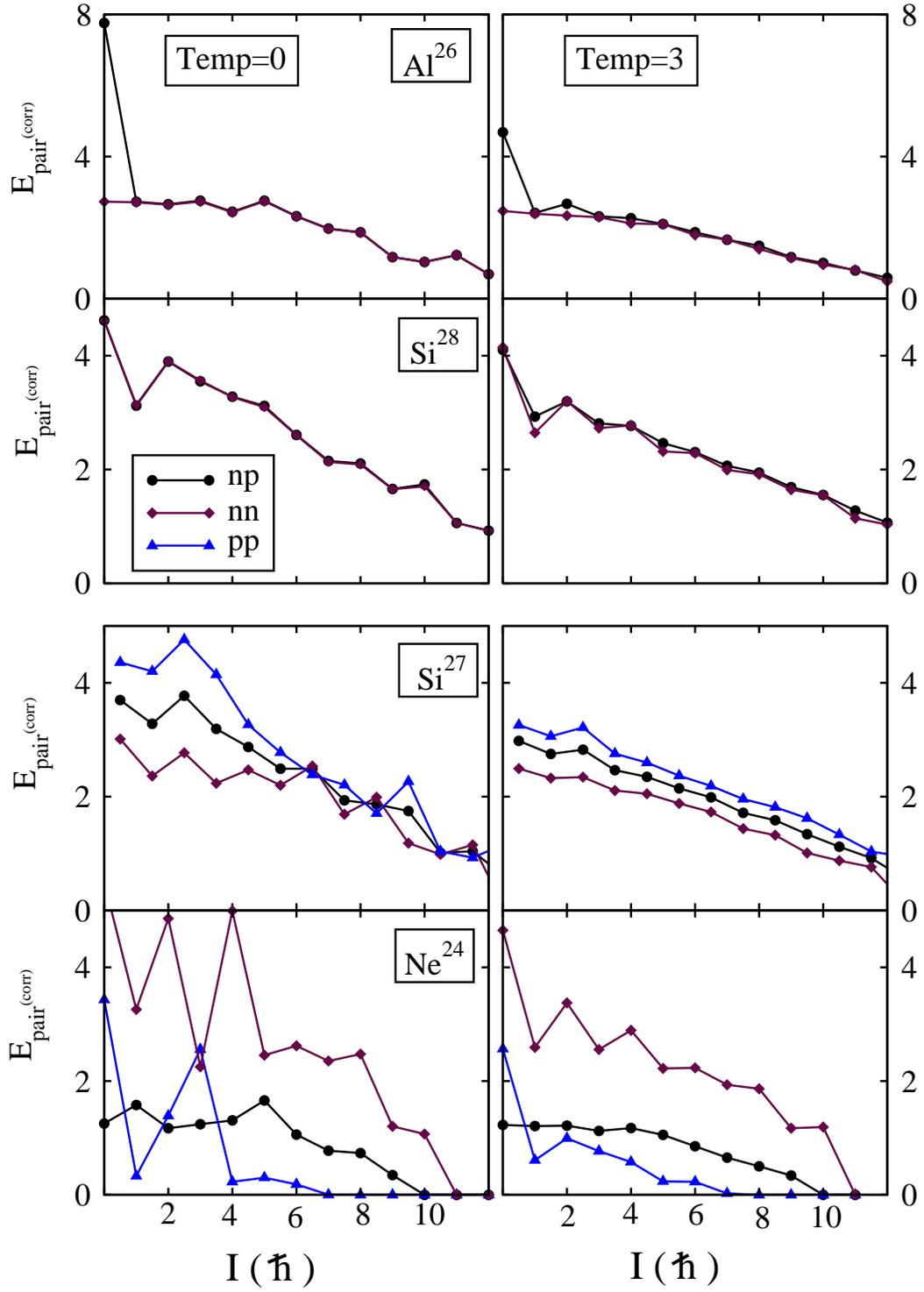}
\caption{(color online) Angular momentum dependence of the pairing energy for $^{26}$Al,$^{28}$Si,$^{27}$Si,and $^{24}$Ne for two different values of temperature.
}
\label{figure.6}
\end{figure*}

\begin{figure*}
\includegraphics {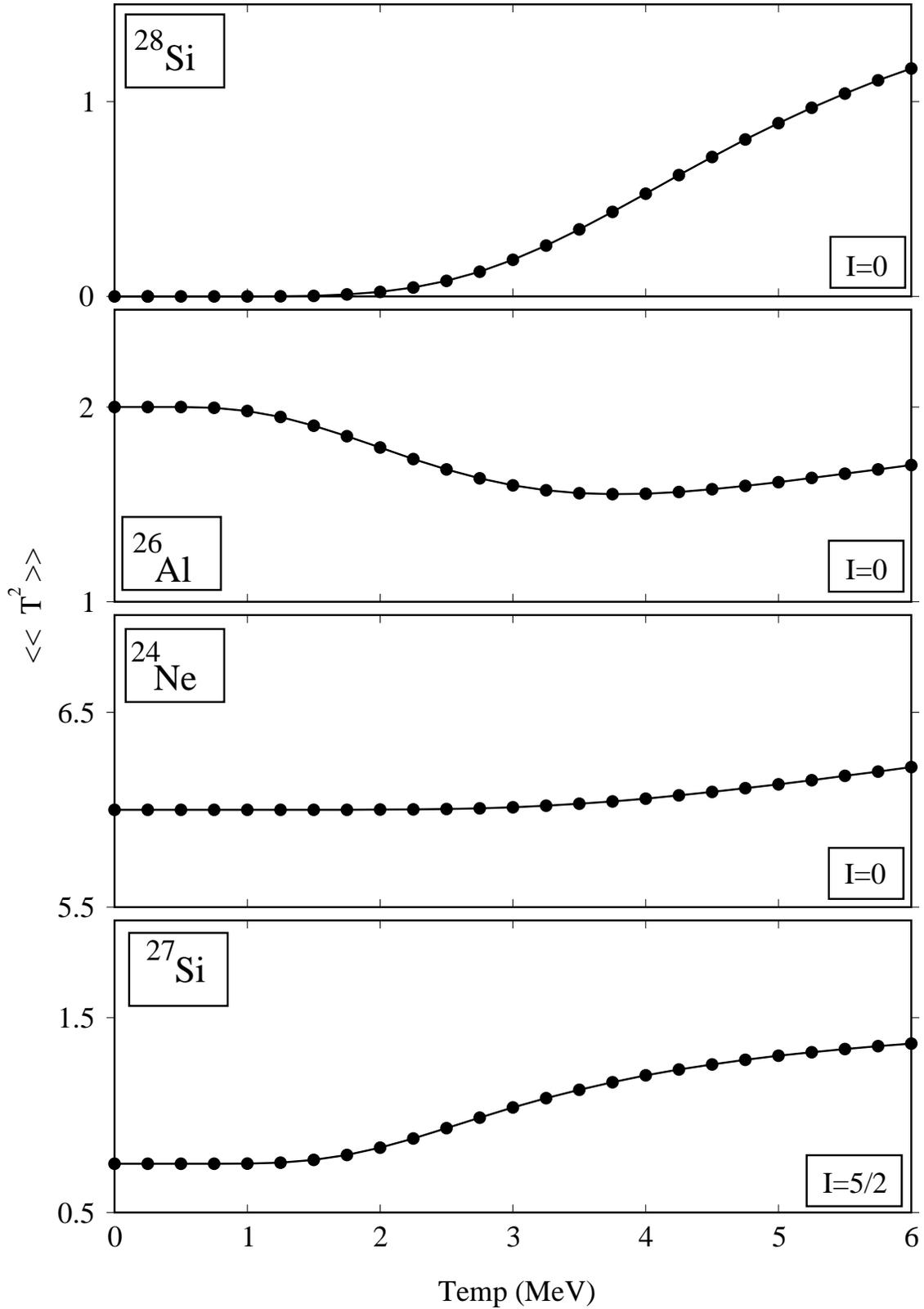}
\caption{(color online) Dependence of average isospin on temperature for the lowest angular momentum ensembles of $^{28}$Si,$^{26}$Al,$^{24}$Ne and $^{27}$Si.
}
\label{figure.7}
\end{figure*}


\begin{thebibliography}{9}

\bibitem{bmp58} A. Bohr, B.R. Mottelson and D. Pines, {\bf 11} (1958) 936

\bibitem{bm75}  A. Bohr and B.R. Mottelson, {\it Nuclear Structure}, Vol.
2 (Benjamin Inc., New York, 1975)

\bibitem{rs80} P. Ring and P. Schuck, The Nuclear Many Body Problem 
(Springer, New York) 1980

\bibitem{good81} A.L. Goodman, Nucl. Phys. {\bf A352} (1981) 30

\bibitem{good88} A.L. Goodman, Phys. Rev.  {\bf C38}  (1988) 1092

\bibitem{er93}
J.L. Egido and P. Ring, J. Phys. G {\bf 19} (1993) 1

\bibitem{drr93}
N. Dinh Dang, P. Ring and R. Rossignoli,
Phys. Rev. {\bf C47} (1993) 606

\bibitem{fkms03}
S. Frauendorf, N.K. Kuzmenko, V.M. Mikhajlov and J.A. Sheikh,
Phys. Rev. {\bf B68} (2003) 024518

\bibitem{aer01}
M. Anguiano, J.L. Egido and L.M. Robledo, 
Nucl. Phys. {\bf A696} (2001) 467

\bibitem{srlr02}
J.A. Sheikh, P. Ring, E. Lopes and R. Rossignoli,
Phys. Rev. {\bf C66} (2002) 044318

\bibitem{landau_stat}
L.D. Landau and E.M. Lifshitz,  {\it Statistical Physics},
Butterworth-Heinemann, (1999)

\bibitem{bfv99}
R. Balian, H. Flocard and M. Veneroni,
Phys. Rep. {\bf 317} (1999) 251

\bibitem{er821}
J.L. Egido and P. Ring, Nucl. Phys. {\bf A383} (1982) 189

\bibitem{er822}
J.L. Egido and P. Ring, Nucl. Phys. {\bf A388} (1982) 19

\bibitem{sr02}
J.A. Sheikh and P. Ring,
Nucl. Phys. {\bf A665} (2000) 71

\bibitem{sdknt07}
M.V. Stoitsov, J. Dobaczewski, R. Kirchner, W. Nazarewicz and J. Terasaki,
Phys. Rev. {\bf C76} (2007) 014308

\bibitem{ee93}
C. Esebbag and J.L. Egido,
Nucl. Phys. {\bf A552} (1993) 205

\bibitem{nt06}
H. Nakada and K. Tanabe, 
Phys. Rev. {\bf C74} (2006) 061301(R)

\bibitem{fo96}
H. Flocard and N. Onishi,
Ann. Phys. (N.Y.) {\bf 254} (1996) 275

\bibitem{spf05}
J.A. Sheikh, R. Palit and S. Frauendorf, 
Phys. Rev. {\bf C72} (2005) 041301(R)

\bibitem{ss06}
J.A. Sheikh and R.P. Singh, to be published

\bibitem{fhmw69}
J.B. French, E.C. Halbert, J.B. McGrory and S.S.M. Wong,
{\it Advances in Nuclear Physics}, edited by
M. Baranger and E. Vogt, Plenum, New York, 1969, Vol.3

\bibitem{cmnpz05}
E. Caurier, G. Martinez-Pinedo, F. Nowacki, A. Poves, A. P. Zuker,
Rev. Mod. Phys. {\bf 77} (2005) 427

\bibitem{usd}
B.H. Wildenthal, Prog. Part. Nucl. Phys. 11 (1984) 5

\bibitem{hz07}
M. Horoi and V. Zelevinsky,
Phys. Rev. {\bf C75} (2007) 054303

\bibitem{dean} D.J. Dean,S. L. Koonin, K. Langanke, P. B. Radha, 
Phys. Lett. {\bf B 399} (1997) 1

\bibitem{engel96}
J. Engel, K. Langanke, P. Vogel,
Phys. Lett. B {\bf 389} (1996) 211

\bibitem{engel98}
J. Engel, K. Langanke, P. Vogel,
Phys. Lett. B {\bf 429} (1998) 215

\bibitem{jaenecke02} J. J\"anecke, T. W. O'Donell, and V. I. Goldanskii,
Phys. Rev. C {\bf 66}, 024327 (2002)

\bibitem{lan06}
K. Langanke, Nucl. Phys. {\bf A778} (2006) 233

\bibitem{bgws92}
C. Baktash, J.D. Garrett, D.F. Winchell and A. Smith,
Phys. Rev. Lett. {\bf 69} (1992) 1500, and references therein

\bibitem{goodman72}
A.L. Goodman,
Nucl. Phys. {\bf A186} (1972) 475

\bibitem{kbo78}
W. Kutschera, B.A. Brown and K. Ogawa, Riv. Nuovo Cim. {\bf 1} (1978) 1

\bibitem{hm72}
J.R. Huizenga and L.G. Moretto, Annu. Rev. Nucl. Sci, {\bf 22} (1972) 427
\end{thebibliography}
\end{document}